# When Is Disclosing Private Conversations with AI to the State Justified? Proactive Referral Policies of Major U.S. AI Companies through the Lens of Contextual Integrity

**Minjong Cheon[1]†, Hanseon Joo[2]†, Ook Lee[2], Jongchang Ahn[2]***

**[1] Department of Computer Science, Korea National Open University, Seoul, Republic of Korea**

**[2] Department of Information Systems, Hanyang University, Seoul, Republic of Korea**

**† These authors contributed equally to this work.**

*** Corresponding author: Jongchang Ahn (ajchang@hanyang.ac.kr)**

**Email: jmj2316@mail.knou.ac.kr; jooking012@hanyang.ac.kr; ooklee@hanyang.ac.kr; ajchang@hanyang.ac.kr**

## Abstract

AI companies engage in "proactive referral," notifying law enforcement of user conversations they judge dangerous without a government request. This poses a dilemma, since referral may expose people who merely confided dangerous thoughts to an AI to investigation, whereas not referring may leave actual violence unprevented. This study asks under what conditions proactive referral can be justified and to what extent companies' public policies disclose those conditions. Drawing on contextual integrity theory and the norms governing exceptional information sharing in confidential counseling, online posts, and private communications, the study develops eight justification criteria classified as required, presumptive, or recommended. Three evaluators analyzed 27 public documents from six major U.S. AI companies using the criteria applicable to corporate policies. No company clarified whether broad sharing provisions or narrower safety notices took precedence, and assessments of referral scope depended on which provisions were given precedence. No company disclosed the specific factual thresholds and exception conditions governing referral decisions, or whether users would be notified afterward. The study turns contextual integrity into a stage-by-stage framework for proactive referral, shows that policy documents must be read in light of their relationships, and proposes policy elements for keeping proactive referral a narrow exception disclosed in advance.

## I. Introduction

People turn to AI not only for searches and work tasks but also with their worries and feelings (Chatterji et al., 2025). Some of these conversations include statements of intent to harm oneself or others, and AI companies can detect them and, if they judge them dangerous, notify law enforcement without a government request, a practice this study calls "proactive referral."

In September 2026, a woman in Florida who wrote to Claude that she would attack the sheriff's office with a gun was arrested on a written-threats charge after Anthropic detected the messages and contacted the authorities (Dellinger, 2026; Krištopaitytė, 2026). Conversely, in June 2025, OpenAI suspended an account used for conversations about gun violence without notifying law enforcement; in February 2026, the account's owner killed eight people in Tumbler Ridge, Canada (Winter, 2026; Supplementary Material S6).

The two cases illustrate the dilemma AI companies face. Keeping conversations with AI in the private sphere may mean failing to prevent violence against others; moving them into the public sphere may lead to the investigation of someone who merely wrote threatening words to an AI and neither prepared nor carried out anything.

In warrants and emergency requests, suspicion comes first; in proactive referral, the company finds conversations by its own criteria and decides whether to report them. Even where the law permits referral, it is the company that first decides which conversations meet the law's requirements, and users can learn its criteria only from public documents.

Prior studies have separately addressed AI companies' privacy policies and corporate reporting obligations, but no study was found that sets out the conditions under which proactive referral of private conversations to state authorities can be justified and reads major companies' policies against those conditions. Treating conversations with AI as a private sphere that should in principle be protected, and proactive referral as an exception to that protection, this study poses two questions.

> ***Research Question 1.*** *Under what conditions can proactive referral, in which a company notifies law enforcement or other authorities of private conversations with AI without a government request, be justified?*
>
> ***Research Question 2.*** *To what extent do major AI companies' public policies disclose those conditions?*

The analysis found that no company stated which of its documents defines the scope of referral, what specific facts and exception conditions decide a referral, or whether users are notified afterward. The second question examines only what public documents disclose, not individual referrals or internal practices.

## II. Background

### 1. The Private and Public Dimensions of AI Conversations

What users write to AI is private, since it is neither addressed to a particular person nor posted for others to read. Users regard chatbot conversations as sensitive (Tran et al., 2025), yet these are stored on company servers where automated systems and reviewers can read them (O6, G1, Me6; document IDs are listed under "Analyzed policy documents" in the references), and when they involve risks to others' life or physical safety, they are not the user's concern alone. The question is therefore when a private conversation may pass into a public flow.

### 2. Definition and Scope of Proactive Referral

In this study, proactive referral means that a company, without a request, identifies signs of a risk to life or physical safety, or of unlawful conduct, in a private user–AI conversation and notifies law enforcement or other state authorities. The definition thus turns on three features, in that the company acts without a request from a state authority, the information comes from a private conversation between a user and an AI, and the recipient is a state authority. Disclosure in response to warrants or emergency requests is therefore not proactive referral, because there the state initiates the flow. Referrals triggered by public posts or tips from other users are excluded. Such referrals begin with content written for others to read or with a third party's report, so they do not raise the question of when a conversation addressed only to an AI may be passed to the state.

Company-initiated disclosure follows either the referral pathway, to law enforcement or government agencies, or the alert pathway, to guardians, a user-designated contact, or crisis support services (Me4, O7). The alert pathway mainly leads to care, whereas through the referral pathway a conversation can become an investigative lead or evidence, so this study analyzes only the referral pathway.

## III. Theoretical Background: Contextual Integrity

### 1. Contextual Integrity and the Information Flows of Proactive Referral

Contextual integrity theory (Nissenbaum, 2004, 2010) holds that information is always shared within a particular context and that each context has norms (context-relative informational norms) that information flows must follow. A privacy violation is judged not by whether information has been made public but by whether a flow conforms to the norms of its original context. An information flow is described by five elements—sender, recipient, information subject, information type, and transmission principle—and because a change in even one of them makes it a different flow, whether that flow is justified must be asked anew. The sender and recipient are the parties between whom information moves, the information subject is the person the information is about, the information type is the kind of information, and the transmission principle is the condition under which it moves, such as confidentiality, consent, or a legal requirement. In a conversation with AI, the user is usually both the sender and the information subject, but other people the user writes about also become subjects (Zhang et al., 2024).

This study divides the flow of proactive referral into the five stages of conversation, detection, review, referral, and subsequent use (Table 1). Once automated systems flag a conversation (detection), staff decide whether to disclose it (review), and at the referral stage it passes to law enforcement, to be used to prevent harm or as evidence (subsequent use). In the Florida case, for example, the messages were flagged by automated monitoring and reviewed by staff, after which law enforcement was notified; the arrest on a written-threats charge marks the point at which the conversation could enter criminal proceedings. What changes across stages is the recipient, the information type, and the transmission principle. At detection, the purpose shifts from generating responses to identifying risks; at review, a person reads the user's writing to decide on referral; at referral, the recipient becomes law enforcement and the information may include account information as well as the conversation; and in subsequent use, the purpose shifts from preventing harm to punishment. Each stage is assessed by what has changed relative to the conversation stage, so that even when each step changes little, it remains visible how far the flow has moved from the original conversation.

**Table 1. Information flows of proactive referral**

| Stage | Sender → recipient | Information type | Transmission principle | What changes |
|---|---|---|---|---|
| Conversation | User → AI model (operated by the company, which retains the conversation) | Thoughts, feelings, questions | To generate responses (the baseline purpose adopted by this study) | Original context (baseline) |
| Detection | Company → automated detection system | User's conversation | To detect policy violations or risks | Processing purpose shifts to risk detection, and a flag is attached |
| Review | Company → company's review staff | Conversation flagged as risky | To decide, under internal criteria, whether to refer | A person reads the user's writing to decide on referral |

| Stage | Sender → recipient | Information type | Transmission principle | What changes |
|---|---|---|---|---|
| Referral | Company → law enforcement | Full or partial conversation, account information | Company's decision criteria within what the law permits (e.g., the SCA emergency exception) | The conversation passes to law enforcement |
| Subsequent use | Law enforcement → prosecutors and courts | Conversation records and account information received | Criminal proceedings | Purpose shifts from harm prevention to punishment |

### 2. Evaluative Framework

Nissenbaum's (2010) decision heuristic identifies what a new flow changes relative to the original context's norms and weighs the effects on values such as autonomy, freedom, and power relations; this study operationalizes it in three ways. First, because norms for AI conversations are not yet established, the baseline is a protection principle under which conversations are used only to provide responses and any other use or external transfer requires justification. This baseline is not a norm observed among users but a normative choice grounded in the basic value of respect for privacy. AI conversations often contain users' private confidences, and companies hold far greater information and power, so this protection matters especially here. Because service safety is also legitimate, detection and review to prevent abuse are accepted if their scope and standards are disclosed, but flows to state authorities must meet strict exception conditions, and the burden of justification lies with the company, which alters the flow and knows the detection standards.

Second, whether passing a user's AI conversation through the company to law enforcement is justified is judged against the laws and professional norms that allow confidentiality to be broken when there is a serious threat to life or safety. Third, the judgment is applied stage by stage, so even if a referral is justified, using the same conversation for punishment is a separate flow that must be justified anew.

### 3. Comparison Contexts

This study compares three contexts that resemble AI conversations and specify when information may be passed on, namely confidential counseling, online posts, and private communications (Table 2). Each resembles AI conversations in one respect and differs in another, in that confidential counseling elicits even dangerous thoughts but involves a professional bound by confidentiality; online posts are stored and monitored by companies but written for others to read; and private communications are retained by a provider but addressed to a person. The norms of confidential counseling and private communications (e.g., the Health Insurance Portability and Accountability Act [HIPAA] and the emergency disclosure exceptions of the Stored Communications Act [SCA] in the United States) require conditions such as a serious risk to life or physical safety, whereas the norms of online posts let companies monitor and remove posts under their own rules and require reporting of some content, such as child sexual abuse material, once discovered (18 U.S.C. § 2258A). Because private AI conversations are not written for others to read and may include help-seeking disclosures whatever task the user came for, this study draws by analogy on the safeguards of counseling and private communications. Companies, however, bear no professional duty of confidentiality, and automated systems read conversations first, so this study requires companies to disclose the decisive factors and procedures in advance and to have a person review each case before referral.

**Table 2. Three comparison contexts**

| Comparison context | Similarities to and differences from AI conversations | Conditions for passing information on |
|---|---|---|
| Confidential counseling (physicians, psychotherapists) | People disclose even dangerous thoughts, as in AI conversations, but AI conversations involve no professional bound by a duty of confidentiality and are typically screened first by automated systems | Confidentiality is the rule. Harm-prevention exceptions generally permit disclosure, to the extent necessary, to those who can prevent the harm when there is a serious (often imminent) risk, in some rules (e.g., Cal. Civ. Code §§ 43.92, 56.10(c)(19)) to a reasonably identifiable or foreseeable victim |
| Online posts (social media, message boards) | Stored on company servers and monitored, but posts are written for others to read | Companies monitor and remove posts under their own rules; like other online service providers, they must report apparent child sexual abuse material to NCMEC once they have actual knowledge of it, without regard to imminence (18 U.S.C. § 2258A) |
| Private communications (email, messaging apps) | Private, with the content retained by the provider, but the counterpart is an AI rather than a person | Confidentiality is the rule. Law enforcement obtains content mainly by warrant (18 U.S.C. § 2703), and voluntary disclosure is limited to statutory exceptions such as consent, emergencies involving danger of death or serious physical injury, or inadvertently obtained contents that appear to pertain to a crime (§ 2702(b)) |

## 4. Prior Research

Contextual integrity has been used as a framework for analyzing privacy policies and chatbot use. Barth et al. (2006) formalized it in logic, incorporating onward flows (a recipient passing information on again) and post-disclosure notification duties into norms, and Shvartzshnaider et al. (2019), in a case study of Facebook's privacy policy, found many information flows that were incomplete or vaguely worded. Tran et al. (2025) found that chatbot users' acceptance of third-party disclosure depends on conditions such as informed consent and de-identification, and Register et al. (2025) identified the problems that arise when an AI therapist's alerting standard is too sensitive or too insensitive but did not specify concrete reporting requirements. Studies applying contextual integrity to large language models have found that models reveal private information in contexts where people would not (Mireshghallah et al., 2024) and that language model agents leak sensitive information even when instructed to protect privacy (Shao et al., 2024). These studies showed that policy documents, AI conversations, and language models' handling of information can be analyzed through contextual integrity, but did not set conditions for flows in which companies hand conversations to law enforcement.

On companies reviewing users' content without a request and alerting outside parties, researchers have warned that on-device scanning to detect child sexual abuse material creates serious security and privacy risks (Abelson et al., 2021) and that automated suicide-risk detection on social media can lead to confrontations with police or unnecessary involuntary hospitalization (Marks, 2019). Platforms in the United States that find child sexual abuse material must report it to the National Center for Missing and Exploited Children (NCMEC), which passes reports on to law enforcement, and a study of this system documented problems with report quality and law enforcement overload (Grossman et al., 2024). Technology companies have also been described as both cooperating with and resisting government surveillance, at times cooperating informally without full legal process ("Developments in the Law—More Data, More Problems," 2018). These studies exposed the risks of company detection and reporting and of companies' cooperation with the state, but did not address the detection of threats against others in private conversations with generative AI and their referral to law enforcement.

Regarding AI companies, King et al. (2025) found that privacy policies often omit key information on the training use and retention of conversations. Lior (2026) proposed that AI companies could owe a narrow, case-by-case duty to protect or warn, modeled on *Tarasoff* (1976), under which a therapist whose patient presents a serious danger of violence to another must use reasonable care to protect the intended victim. De Mello Barreto (2025) proposed safeguards such as data minimization, purpose limitation, legal process, and independent oversight for legally mandated duties to monitor and report AI misuse. Christakis (2026) identified police referral by providers as one way in which chatbot conversations are exposed, noting that companies make these decisions privately and at scale under an emergency disclosure framework built for email, with criteria, volumes, and error rates largely undisclosed. These studies proposed norms for AI companies or examined the legal status of conversations with them, but none set out the conditions that justify referral stage by stage and then compared companies' current policies against them.

Following Barth et al. (2006), who incorporated onward flows into norms, this study treats proactive referral as a chain of information flows running through detection, review, referral, and subsequent use, derives justification criteria for each stage, and uses them to compare the public policies of six major U.S. AI companies.

## IV. Justification Criteria for Proactive Referral

### 1. Method of Derivation

Drawing on the exception norms of the three comparison contexts and the conditions and purposes of U.S. law, this study constructed eight criteria for justifying a change in the elements of an information flow. For private communications, without resolving the SCA's direct applicability to AI chatbot providers, this study used its emergency disclosure exceptions as a comparative benchmark. These exceptions resemble proactive referral aimed at preventing harm in purpose and structure, as they specify the conditions under which providers may voluntarily disclose stored information to government authorities to prevent death or serious physical injury (18 U.S.C. § 2702(b)(8), (c)(4)). The SCA also permits providers to disclose information to protect their own rights or property (18 U.S.C. § 2702(b)(5), (c)(3)), but this exception was not used as a benchmark because it protects the provider's interests rather than preventing harm to persons. The eight criteria are designed to ask, at least once, whether each change in recipient, information type, and transmission principle at each stage is justified, and notice is assessed across all stages.

### 2. Eight Criteria

The eight criteria cover conditions for referral (Criteria 1–4), manner of referral (Criteria 5 and 6), and subsequent use and notice (Criteria 7 and 8) (Table 3). Each criterion has a strength level, where "required" denotes a condition indispensable to justification; "presumptive," one that must be met except under narrow exceptions disclosed in advance; and "recommended," one that is not necessary for justification but strengthens protection when met. The levels thus differ in what follows when a condition is not met, in that without a required condition a referral cannot be justified, without a presumptive condition it can be justified only under an exception disclosed in advance, and without a recommended condition it may be justified but protects users less. Strength reflects how consistently a condition is required by three harm-prevention norms, which are HIPAA's disclosures to avert a serious threat, the SCA's emergency disclosure, and California's rules on a psychotherapist's disclosure to prevent a serious and imminent threat and on the duty to protect (Cal. Civ. Code §§ 56.10(c)(19), 43.92). Agreement among these norms is a comparative benchmark rather than what settles strength by itself, because strength also reflects the seriousness of the harm, the gap in information and power between companies and users, and the hard-to-reverse consequences of passing information to the state. A condition on which all three norms agree expresses a shared judgment about when confidential information may be passed on, whereas a condition found in only some of them reflects choices that differ across settings and can therefore yield to a narrow exception. Serious harm and a judgment grounded in facts or good faith, demanded by all three, are required (Criterion 1 and Criterion 4(i)), and an identifiable target, demanded only by the California rules, is presumptive (Criterion 3). HIPAA (45 C.F.R. § 164.512(j)(1)(i)) and California's disclosure rule (Cal. Civ. Code § 56.10(c)(19)) expressly require a serious and imminent threat, and the SCA requires an emergency that calls for disclosure without delay, but this study nonetheless treats imminence as presumptive (Criterion 2), because requiring it without exception would miss conversations that reveal concrete preparations for an attack but leave its timing unclear. Information minimization and advance notice are required as general principles of data protection law and because users must be able to anticipate and contest significant flows to state authorities (Criteria 6 and 8). The condition that none of the three norms requires and that this study adds (proactive notification after referral) and the condition the state must meet (Criterion 7) are recommended, while human review (ii) and advance disclosure of decisive factors (iii) under Criterion 4 are required.

**Table 3. Justification criteria for proactive referral and their bases**

| Criterion | Question | Stage and element | Strength | Basis | Sources |
|---|---|---|---|---|---|
| 1. Severity | Does it involve serious harm to life or physical safety? | Referral: transmission principle | Required | Legal analogy | U.S.: SCA (§ 2702, death or serious physical injury), HIPAA (serious and imminent threat to health or safety), California law (§ 56.10(c)(19), serious and imminent threat; § 43.92, serious threat of physical violence) |
| 2. Imminence | Is the harm about to occur? | Referral: transmission principle | Presumptive (narrow exceptions) | Legal analogy | U.S.: SCA (emergency requiring disclosure without delay), HIPAA (imminent threat), California law (§ 56.10(c)(19), imminent threat). The narrow exception is this study's proposal |
| 3. Identifiable target | Is it apparent who or what (person, place, group) is at risk? | Referral: transmission principle | Presumptive (narrow exceptions) | Legal analogy | U.S.: California law (Cal. Civ. Code § 43.92, reasonably identifiable victim; § 56.10(c)(19), reasonably foreseeable victim); not required by HIPAA or the SCA |

| Criterion | Question | Stage and element | Strength | Basis | Sources |
|---|---|---|---|---|---|
| 4. Judgment: who decides, and on what basis | (i) Is the decision based on facts revealed in the conversation? (ii) Does a human review before referral? (iii) Are the decisive factors (e.g., target, means, timing) disclosed in advance? | Review: recipient (who); referral: transmission principle (on what basis) | Required | (i) Legal analogy; (ii) and (iii) Proposed by this study | (i) U.S.: HIPAA (good-faith belief, presumed when based on actual knowledge or a credible representation, § 164.512(j)), SCA (good-faith belief), California law (§ 43.92, serious threat actually communicated by the patient). (ii) cf. GDPR Art. 22 (human intervention in automated decisions). (ii) and (iii) are this study's proposals |
| 5. Appropriate recipient | Is it sent to a recipient suited to the risk (only risks to others go to law enforcement; self-harm does not)? | Referral: recipient | Recommended | Proposed by this study | U.S.: HIPAA (disclosure to persons reasonably able to prevent or lessen the threat, including the target). Excluding self-harm is this study's proposal |
| 6. Information minimization | Are the types of information transferred disclosed, and is only what is necessary transferred? | Referral: information type | Required | Legal analogy | U.S.: HIPAA (minimum necessary), SCA (only communications relating to the emergency) |
| 7. Separate control over subsequent use (state) | Is use of the referred conversation itself as grounds for punishment separately controlled? | Subsequent use: recipient, transmission principle | Recommended (state) | Legal analogy | U.S.: rules on substance use disorder treatment records (42 U.S.C. § 290dd-2(c): records may not be used against the patient in criminal or other proceedings without a court order or consent) |
| 8. Advance notice and post-referral notification | Are criteria and procedures disclosed in advance, and is the user informed of the referral after the risk has ended? | Detection, review, referral: transmission principle (advance notice); after referral (post-referral notification) | Advance notice: Required; post-referral notification: Recommended | Advance notice: Legal analogy; post-referral notification: Proposed by this study | U.S.: HIPAA (notice of privacy practices, § 164.520; accounting of disclosures on request, § 164.528, which covers § 164.512(j) disclosures). Proactive notification after referral is this study's proposal |

*Note.* The provisions are comparative bases for the criteria, not obligations of AI companies under current law.

**2.1 Conditions for Referral (Criteria 1–4)**

Serious harm to life or physical safety is required (Criterion 1), and imminence and an identifiable target are presumptive (Criteria 2 and 3), but a narrow exception under requirements disclosed in advance is permitted where means and preparations are concretely evident and a human has confirmed an intent to attack and a realistic possibility of carrying it out, even if the target or timing is unclear. Severity is required because all three harm-prevention norms reserve disclosure for serious harm, so a risk to property or rights alone does not justify referral. When a risk is imminent and its target is identifiable, a referral is more likely to actually prevent harm and less likely to rest on mere speculation. Requiring both without exception, however, would rule out referring a conversation that concretely reveals preparations for an attack, such as obtaining a weapon, even when no date or target has been set. For imminence, which all three norms require in some form, this exception is therefore this study's normative proposal. Criterion 4 comprises (i) a good-faith judgment grounded in facts evident in the conversation, (ii) human review before referral, and (iii) advance disclosure of the factors that decide a referral (facts such as target, means, timing, and preparation) and of the exceptions; (i) is borrowed from the three harm-prevention norms, whereas (ii) and (iii) are conditions newly proposed by this study. Elements (ii) and (iii) are required because human review adds a check on automated judgments, which can err, and makes a person responsible for the decision; because referral can lead to hard-to-reverse outcomes such as arrest; and because users must be able to anticipate and contest the judgment. Companies must disclose in advance the factors that decide a referral and the conditions for exceptions, rather than the detection algorithm.

**2.2 Manner of Referral (Criteria 5 and 6)**

This study recommends that referral to law enforcement be limited to risks to others and that risks of self-harm be routed instead to crisis support or emergency services that can actually reduce the risk (Criterion 5). Suicide and self-harm are not crimes in most jurisdictions (Mishara & Weisstub, 2016), fear of coercive intervention can lead people to conceal suicidal thoughts from clinicians (Richards et al., 2019), and such intervention can deter help-seeking (Saltz & Leibowicz, 2026). Because HIPAA also permits disclosing a risk of self-harm to the police, this criterion is a proposal of this study and is therefore recommended rather than required. Companies should also specify the types of information they hand over and limit the information to what is necessary to prevent the risk (Criterion 6). Minimization matters because a referral can include the full conversation and account information, exposing private matters unrelated to the risk.

**2.3 Subsequent Use and Notice (Criteria 7 and 8)**

Even when a referral is justified, using the referred conversation to punish the user for that conversation itself requires separate controls (Criterion 7). A flow whose purpose and recipient change must be justified anew, and some threat statutes presuppose that a threat was communicated to, or could be viewed by, another person, so whether text written only to an AI meets that requirement calls for a separate legal judgment. Because the state must establish this condition through legislation and criminal procedure, it is excluded from the evaluation of companies.

So that users can anticipate significant flows to state authorities and judge and contest their appropriateness, companies are required to give advance notice of the criteria and procedures for detection, review, and referral and are encouraged to inform the user of a referral once the risk has passed (Criterion 8). Post-referral notification is only recommended because no norm examined in this study requires notifying the user proactively (HIPAA provides only an accounting of disclosures on request) and because notifying the user while the risk persists could defeat the purpose of referral.

### 3. When the Law Provides for Referral

Because the SCA merely permits voluntary disclosure in emergencies, the conditions and procedures U.S. companies set for themselves are evaluated against Criteria 1–6 and 8. Where a statute instead mandates reporting, as for child sexual abuse material (18 U.S.C. § 2258A), whether reporting is confined to serious, imminent risks depends on the statute rather than on company policy, so such reporting falls outside this evaluation.

## V. Methods

### 1. Research Design

This study is an interpretive document analysis (Bowen, 2009) of 27 public documents from six U.S. AI companies. Three evaluators, all of whom are authors of this study, reviewed all of each company's documents as a set, following the information flow elements (Shvartzshnaider et al., 2019) and the criteria in Section IV, and judged each criterion on the basis of the text and its context. OpenAI's letter to the Canadian government and the reported incidents (Supplementary Material S6) served only as auxiliary sources explaining the background and changes in the criteria.

### 2. Companies Analyzed

The analysis covers six U.S. companies that develop their own large language models, operate general-purpose consumer chatbots, and publish third-party disclosure policies, namely OpenAI (ChatGPT), Anthropic (Claude), Google (Gemini), Microsoft (Copilot), Meta (Meta AI), and xAI (Grok) (Supplementary Table S6-5). The analysis is limited to U.S. companies for two reasons. U.S.-based institutions produced the most notable AI models in 2024 (40, compared with 15 from China and 3 from Europe; Maslej et al., 2025), and ChatGPT alone had been adopted by around 10% of the world's adult population by July 2025 (Chatterji et al., 2025). In addition, because U.S. law permits but does not require emergency disclosure (Section IV.3), differences among companies under the same legal structure reflect the companies' own choices.

### 3. Document Collection

In October 2026, privacy policies, terms of use, safety notices, government request policies, and government request reports were collected from the companies' own pages (access dates in Supplementary Table S6-3); request documents were included because a company that disclosed proactive referral would likely do so there. Of these, 27 documents used to judge or compare the conditions of the referral pathway and their disclosure were analyzed, including alert-pathway documents used only for comparison (O7, Me4). Six collected items, such as earlier versions and a news report, were excluded, as were documents without referral conditions, and the U.S. version was used for OpenAI (Supplementary Tables S2-2 and S2-3).

### 4. Reading Principles

Because a company's documents differ in timing and in the conditions they state, the following four principles were applied to all companies.

Principle 1. All of a company's documents are read together, so a condition stated in any of them counts as disclosed. Where documents of the same type state the same matter differently, the most recent governs.

Principle 2. Where a document or provision deals specifically with referral to law enforcement or unrequested disclosure (a referral-specific provision), the referral-specific provision determines when referral occurs, and general provisions on sharing personal information are used only to supply what the referral-specific provision does not address. Where a company maintains several distinct referral pathways, each pathway is read separately and the results are reflected together in the company's evaluation.

Principle 3. A sentence is read together with the heading under which it appears and the sentences around it. A sentence under the heading "Requests for information in emergencies" is read as disclosure in response to requests (G2), and situations listed with "such as" or "etc." are not taken to narrow the scope of referral to those situations.

Principle 4. A sentence that addresses only disclosure in response to government requests is not treated as a basis for proactive referral. A sentence that addresses unrequested disclosure is read as a basis even if it appears in a request-handling document (e.g., the Q&A in M1 on what happens "absent a formal legal request").

A provision read as a basis for one criterion was applied in the same way to every other criterion it addresses. Judgments rested solely on what the documents state and were not supplemented with news reports or other outside sources.

### 5. Interpretive Transparency

Because companies' documents do not state referral conditions in a fixed format, assigning sentences to judgment categories involves interpretation. For each criterion, categories are ordered from the most specific disclosure to no statement of the condition, and element-level judgments for the multi-element Criteria 4, 6, and 8 are reported in Supplementary Table S4-4. For severity, a company was placed in one of four categories according to whether it narrowed referral to risks to life or physical safety (including violence), mentioned risks to life or physical safety as grounds for referral but did not narrow referral to them, described risk broadly, e.g., as "safety" or "harm" (listing property and rights together), or did not state the type of risk. Three evaluators (A, B, and C), given the scope rules, reading principles, category guidelines, key passages for each company with their translations, and the full source texts, independently judged 48 cells (six companies × Criteria 1–6, advance notice, and post-referral notification). The scope rules set aside disclosure in response to government requests, the alert pathway, statutory reporting of child sexual abuse material, and referrals triggered by public posts or tips. After the principle on priority among documents was settled, the three evaluators re-judged only the 7 cells it affected (Criteria 1–3 for OpenAI; Criteria 1–3 and 6 for Microsoft), and all other cells kept their first-round judgments. In the first round, before the re-judgment, the three evaluators agreed on 40 of 48 cells (83%), and in the final judgments on 41 (85%); all split cells were 2–1, and the final judgments followed the majority. In this study, "did not state" means that the condition was absent from the documents found.

In the final judgments, Evaluators A and C agreed on 44 cells and the other two pairs on 43 each, and no cell was read differently by all three. All three agreed for every company on imminence, identifiable target, judgment, recipient, and post-referral notification, but for only four companies on severity and advance notice, and three on information minimization.

## VI. Results: How Far Do Companies' Policies Disclose the Criteria?

Four companies stated that they may make referrals or reports to state authorities, whereas xAI and Google had only disclosure and sharing provisions that do not specify recipients. Table 4 summarizes each company's supporting wording, with document IDs in parentheses (Supplementary Material S6).

**Table 4. What companies' documents state about referral pathways (source texts in Supplementary Material S1)**

| Company | Risks referred (Criteria 1–3) | Judgment (Criterion 4) | Recipient (Criterion 5) | Information transferred (Criterion 6) | Advance notice (Criterion 8) |
|---|---|---|---|---|---|
| OpenAI | Imminent and credible risk of harm to others; planned violence (O6). General provision: sharing with government for the safety of users and the public, rights and property, and in connection with violations of law (O5) | Review by trained personnel of conversation content, surrounding conversation, and behavior patterns (O6) | Law enforcement (O1, O6). Self-harm cases not referred to law enforcement as of August 2025 (O1) | Personal information, including information on use of the services (O5) | Detection → human review → in-depth investigation → referral (O6) |
| Meta | Among the seven situations: "emergency situations, such as risk of death or imminent bodily harm," protection of rights and property, fraud prevention, etc.; protection of users or others (Me3) | Good-faith belief (Me3). Review may be automated or human, not linked to referral (Me6) | Law enforcement, government authorities, affiliates, etc. (Me3) | Information (Me3) | Situations; link between detection, assessment, and contact with law enforcement (Me3, Me5). Decision-maker not specified |
| Microsoft | Without a request: crimes against the company, its employees, or its customers; emergencies involving death or serious physical injury (M1). General provision: safety of customers, organizations, and the public; rights or property of Microsoft, including referral of stolen-property trafficking to law enforcement (M3) | Belief (M1); good-faith belief (M3) | Law enforcement (M1) | Limited information (M1); personal information, including content such as emails and OneDrive files (M3) | Situations only (M1) |
| Anthropic | Serious harm to any person or to property (A6); sole discretion (A7) | Good-faith belief (A6) | Government, law enforcement, etc. (A6, A7) | Information including but not limited to inputs, outputs, or actions (A7) | Situations only (A6) |
| xAI | Protection of the rights and property of the company, its agents, customers, and others, and of the personal safety of the company, its customers, or any person (X2) | Not stated; personnel may review content, not linked to referral (X3) | Not stated | User content (inputs and outputs) (X2) | Purpose only (X2) |
| Google | Protection against harm to the rights, property, or safety of Google, its users, or the public (G4) | Good-faith belief (G4); human reviewers may read chats, not linked to referral (G1) | Not stated (outside Google) | Personal information (G4) | Purpose only (G4) |

## 1. Conditions for Referral (Criteria 1–4)

### 1.1 Severity

OpenAI was the only company that narrowed referral to risks to life or physical safety. Its 2026 notice (O6) named "an imminent and credible risk of harm to others," that is, "planned violence," as the ground for referral, but because its later privacy policy (O5) retained sharing with governments to protect rights and property, among other purposes, this result rests on Principle 2 (Section V.4), under which the notice, not the privacy policy, decides when referral occurs. Microsoft listed emergencies as a separate ground, but the same Q&A also provides for reporting crimes against the company, its employees, or its customers, leaving room for referral even without a risk to life or physical safety (M1). Microsoft's Q&A, for example, states that "If one of our customers or employees, or Microsoft itself, is the victim of a crime, we may report some limited information to law enforcement" (M1). The others bundled life or physical safety with rights or property (Anthropic, and Meta, whose sharing provision lists emergencies alongside protecting rights and property and preventing fraud; Me3) or mentioned only "safety" and "rights and property" (xAI and Google).

### 1.2 Imminence

OpenAI was also the only company that narrowed referral to imminent risks (O1, O6). Meta and Microsoft cited imminent or emergency situations as one ground for referral but allowed referral in other situations as well, and the other three companies did not mention imminence as a condition in the provisions this study counted as grounds for referral. In the sharing provision of Meta's privacy policy, "imminent bodily harm" appears only as an example within one of seven listed situations in which information is shared (Me3). Anthropic's reference to "imminent physical harm or death" appears in an exception within its government request policy and was therefore not counted (A1).

### 1.3 Identifiable Target

No company stated that the target of the risk must be identifiable. OpenAI narrowed referral to risks to others but said it may refer even when the user has not explicitly stated the target, means, or timing (O6); this means only that the target need not be stated, and the documents do not say whether it must be identifiable from context. The rest described the target broadly, using phrases such as "you or others," "any person," and "the public." xAI, for example, may disclose information "to protect the personal safety of SpaceXAI, its customers, or any person" (X2), wording that names whose safety the company may protect, not whom a conversation must threaten before it is referred.

### 1.4 Judgment

Only OpenAI stated that humans decide whether to refer. By stating that trained staff review the conversation, surrounding exchanges, and behavior patterns (O6), OpenAI disclosed fact-based judgment (i) and human review (ii), but "imminent and credible" describes only a degree of risk, not which facts decide a referral (iii). Meta, xAI, and Google stated that humans may view conversations but did not link this to referral (Me6, X3, G1), and Anthropic and Microsoft stated only a degree of belief, such as a "good-faith belief" (A6) or "where we believe" (M1). Meta's AI terms, for instance, state that review of interactions with AIs "may be automated or manual (human)" (Me6), which means that a person may see a conversation, not that a person decides whether to refer it.

## 2. Manner of Referral (Criteria 5 and 6)

### 2.1 Recipient

Four companies identified a recipient, but OpenAI alone restricted law enforcement referrals to risks to others (O1, O6), stating in its August 2025 notice that it was "currently not referring self-harm cases to law enforcement" (O1). The other three named law enforcement or government agencies as recipients without limiting law enforcement referral to risks to others, while xAI and Google stated only that they "disclose" information and "share personal information outside of Google," respectively. For these five companies, the documents therefore do not rule out that a conversation revealing a risk of self-harm may also be sent to law enforcement.

### 2.2 Information Minimization

Only Microsoft identified the types of information passed on and stated that it limits the amount, citing personal data "including your content (like the content of your emails in Outlook.com or files in OneDrive)" (M3) and "limited information" (M1), though "limited information" is not the same as the minimum necessary. OpenAI, Anthropic, and xAI identified only the types of information (O5, A7, X2), and Meta and Google referred only to "information" or "personal information." Anthropic, for example, reserves the right to report information "including but not limited to Inputs, Outputs, or Actions" (A7), naming types while leaving the scope open.

## 3. Advance Notice and Post-Referral Notification (Criterion 8)

### 3.1 Advance Notice

Only OpenAI disclosed both the procedure from detection to referral and its referral criteria, describing automated detection, contextual review by trained staff, and in-depth investigation based on structured criteria, followed by notification of law enforcement when the risk is imminent and credible (O6). Its notice states that a case presenting "indicators of potentially serious, real-world harm" is escalated for a more in-depth investigation, "including assessing the overall level of risk using structured criteria" (O6), but the documents found do not say what those criteria are. By stating that it has systems to detect harmful conduct and may contact law enforcement "based on our assessment," Meta disclosed that detection can lead to referral (Me3, Me5) but not who makes the assessment. Microsoft and Anthropic disclosed only the situations in which they refer, and xAI and Google only their purposes.

### 3.2 Post-Referral Notification

No company stated that it notifies users after a referral, although five companies set out a principle of notifying users of government requests for their information (O2, A1, M1, Me1, G2), and Google and Meta stated that notice delayed by an emergency is given once the reason no longer applies (G2, Me1). Google, for example, states that "we'll provide notice if we learn that the emergency has passed" (G2), but no company applied this practice to referrals it initiates itself.

## 4. Alternative Readings

The reading that could most alter the results concerns the priority among documents. Because the companies did not state which provisions define the scope of referral, Principle 2, which gives precedence to referral-specific provisions, is an analytical choice of this study, not a precedence the companies themselves state; the three evaluators initially read the broad general provisions as the scope the companies had retained (Supplementary Materials S4 and S5). As Supplementary Table S6-1 shows, when referral-specific provisions take precedence, OpenAI counts as having narrowed referral and Microsoft as having limited the information passed on; when general provisions take precedence, no company does.

Whichever reading is adopted, no company requires that the target of the risk be identifiable, specifies what facts must emerge for a referral, or states that it notifies users after a referral. Three other readings that could change the judgments, including the scope of Microsoft's Q&A (M1), are described with their effects in Supplementary Material S4.

### 5. Differences across Jurisdictions and over Time

No analyzed document stated that different referral criteria apply depending on the user's jurisdiction. Changes in criteria surfaced first outside the policy documents, as when OpenAI told the Canadian government after the Tumbler Ridge shooting that it had made its criteria more flexible (OpenAI, 2026), and its April 2026 notice (O6) added a criterion under which the target, means, or timing need not be explicitly stated, but this criterion was not carried over into its subsequently updated privacy policy (O5).

## VII. Discussion

### 1. Blurred Boundaries between Narrow Notices and Broad Sharing Provisions

The most salient finding of the analysis is that the boundary of referral remains undefined in the gap between narrow safety notices and broad sharing provisions. OpenAI and Microsoft set out narrow criteria or limits on information in safety notices or Q&As, but left the broad sharing provisions in their privacy policies unchanged. If the narrow criteria are the rules actually applied, the companies need only add to the broad provisions an exception stating that proactive referral of conversations follows the criteria of the safety notice, so that every document shows the same limits. Two of the seven cells on which the final readings diverged (Microsoft's severity and information minimization) likewise involved a company that described the scope of referral differently across documents or provisions (Supplementary Material S4). A user who reads only OpenAI's safety notice would conclude that only risks of harm to others are referred (O6), whereas one who reads only its privacy policy would find sharing with governments to protect rights and property, among other purposes (O5).

The relationship between documents is not the only factor that blurs the boundary. Even within a single document, companies gave examples instead of conditions ("emergency situations, such as risk of death or imminent bodily harm," Me3) or stated purposes and beliefs instead of situations ("to protect safety," "when we have a good faith belief"), leaving themselves discretion over which conversations to refer without clearly specifying the applicable conditions.

There are presumably reasons for leaving the boundary open. Risks cannot all be specified in advance, narrow criteria may miss real attacks, and companies want room to avoid liability, but because the burden of justification lies with the party departing from the baseline of protection (Section III.2), here the company that sends conversations outside, the cost of a blurred boundary should fall on companies. OpenAI and Meta specified covered conversations and reviewers for their alert pathways (O7, Me4), so comparable specificity is possible for the referral pathway. If users cannot know the criteria, even those seeking help may say less. The evidence is analogical and comes from government surveillance; after the Snowden revelations, views of privacy-sensitive Wikipedia articles fell (Penney, 2016), searches for terms that users thought could draw government scrutiny declined (Marthews & Tucker, 2014), and most respondents in a survey said that knowing the government monitors online activity would make them more careful about what they search for (Penney, 2017). Users of chatbots based on large language models already weigh privacy risks against benefits when deciding what to disclose (Zhang et al., 2024), so uncertainty about referral may also shape what they say. Disclosing specific criteria is therefore a condition of justification and may reduce chilling effects.

### 2. The Cost of the Imminence Criterion and a Narrow Exception

The imminence criterion comes at a cost. In the Tumbler Ridge case, OpenAI made no referral, stating that it had found no credible and imminent plan, and in its February 2026 letter described criteria it had already made more flexible, so that it can refer even when the user has not explicitly stated the target, means, or timing (OpenAI, 2026). Because the account had already passed through detection and human review before being suspended, the case turned not on whether such conversations are found but on where the threshold for referral is set (Supplementary Material S6). This change reduces the chance of missing attacks but may extend referral into the realm of prediction, referring users for what they might do rather than for plans their conversations reveal, so this study proposes reserving referral to law enforcement, as a rule, for imminent risks with an identifiable target and handling other concerns through user-directed measures such as warnings, account restrictions, and pointing users to counseling resources.

Some risks, however, cannot be prevented by user-directed measures alone. If a conversation concretely reveals preparations for an attack, such as obtaining a weapon or scouting a location, without naming a target or date, the plan may continue elsewhere even if the account is blocked, and in the Tumbler Ridge case suspending the account did not prevent the attack. This is why Section IV allows a narrow exception. Conversely, if fiction, jokes, or emotional venting are read as threats, people who have done nothing may face investigation and users may hold back even when they need help, so defining and publishing in advance which facts lead to referral may reduce both missed attacks and erroneous referrals (Supplementary Table S6-2).

### 3. A Gateway between Referral and Punishment

No company stated that it restricts how referred conversations may be used, and by citing “investigating suspicious activity” (Me3), Meta presupposed use in investigations. A company can limit what it hands over, but once a conversation reaches law enforcement, its use is decided by investigators, prosecutors, and courts, and at that stage the purpose of the flow shifts from preventing harm to punishment. The gateway called for by Criterion 7 must be built by the state; as one possible mechanism modeled on the U.S. restriction on criminal use of substance use disorder treatment records (42 U.S.C. § 290dd-2(c)), this study proposes requiring separate judicial review before a referred conversation is used to punish the user for the conversation itself, while its use as evidence of other crimes would follow ordinary procedures. This is not a proposal to immunize AI conversations but to control separately the purpose, scope, and procedure of their criminal use, and the treatment-records rule serves only as an analogy, since its protective purpose and relationships differ from those of general-purpose AI conversations. The review could examine whether the referral met Criteria 1–4 and whether there was conduct beyond the conversation, such as preparation or execution. Such review would separate two questions that a single referral otherwise merges, asking separately whether passing on the conversation was justified to prevent harm and whether the conversation itself may ground punishment.

### 4. Policy Recommendations

Table 5 summarizes the discussion as proposals that take the protection of users’ private conversations as the principle and treat referral to law enforcement as a narrow exception published in advance.

**Table 5. Policy Recommendations**

| Proposal (corresponding criteria) | Description | Responsible party |
|---|---|---|
| 1. Referral requirements stated as conditions (1–4) | State requirements such as "only when the conversation shows an imminent risk of death or serious physical injury to an identifiable person, place, or group" (subject to a narrow exception disclosed in advance), and disclose the facts used in the decision (target, means, timing) and that a human decides before referral | Companies; supervisory authority guidance (in jurisdictions where referral is mandatory, statutory requirements apply) |
| 2. Consistency across documents (1–6, 8) | Add to broad sharing and discretion provisions an exception that "proactive referral of conversations follows the criteria of the safety notice," or state which document prevails | Companies |
| 3. Separation of recipients (5) | Limit law enforcement referrals to imminent risks to others, do not refer self-harm, and handle other concerns through user-directed measures; disclose any requirements for exceptional referrals in advance | Companies, supervisory authorities |
| 4. Information minimization (6) | Transfer only the minimum information needed to prevent the risk (such as relevant excerpts and necessary identifying information), disclose its types, and require law enforcement to seek further information through legal process | Companies, data protection authorities |
| 5. Gateway for criminal use (7) | Require separate judicial review before a transferred conversation is used to punish the user for the conversation itself. Use as evidence of other crimes follows ordinary procedures | Legislatures |
| 6. Disclosure of procedures and jurisdiction-specific criteria (8) | Disclose how detection, review, and referral are linked, who decides, and whether criteria differ by jurisdiction | Companies, national supervisory authorities |
| 7. Post-referral notification (8) | Inform the user after the risk has ended; delay or omit notice where it would obstruct an investigation, increase the risk, or breach legal confidentiality, and record the reasons | Companies, supervisory authorities |

## 5. Theoretical Implications

First, extending work on onward transmission (Barth et al., 2006) and alerting standards (Register et al., 2025) to flows toward state authorities, the study set graded criteria for each stage and proposed, as Criterion 7 (separate control over subsequent use), that justifying a referral does not justify criminal use. Second, because whether a transmission principle actually limits a flow depends on what is written, where, and how, and a company may state narrow and broad conditions in different documents, policy documents must be read not only provision by provision but also in light of their relationships. Third, prior research has viewed companies as surveillance intermediaries that screen government requests (Rozenshtein, 2018), but in proactive referral companies generate suspicion within private conversations. Because companies set operational referral thresholds in their own documents within applicable legal limits, the practice connects with private governance in the shadow of the state (Bloch-Wehba, 2019).

## 6. Limitations and Future Research

First, the analysis covered public documents, which may differ from internal practice, and is a snapshot of the collection period. Second, the documents are uneven, because the law enforcement documents of Google, Microsoft, and Meta apply company-wide, Grok within X was not reviewed, and no law enforcement policy was found for xAI. Third, because the three evaluators followed the same guidelines, their agreement shows only how consistently the guidelines were applied, not that the guidelines themselves are sound; agreement also varied by criterion, being lower for severity, advance notice, and information minimization, and the principle on priority among documents in particular changes the results for OpenAI and Microsoft (Section VI.4). Fourth, the analysis covered only U.S. companies and U.S. statutory text, and the choice of criteria involves normative judgment. Future research could trace earlier versions of policies to see how AI companies' referral criteria change in response to incidents or regulation, and could use transparency reports or known cases to check whether the published criteria match actual referrals. User surveys could examine whether users share the expectation that conversations with AI are private and whether awareness of referral policies reduces help-seeking speech. The comparison could also be extended beyond the United States to how jurisdictions such as the EU, China, and Korea treat proactive referral by AI companies.

## VIII. Conclusion

In conclusion, because private conversations with AI also contain thoughts and feelings that never progress to preparation or execution, the principle, in light of modern criminal law's refusal to punish thought alone, is protection. Whether such text can be punished is a separate question for criminal law, but whether a company hands it to the state must start from this principle. Yet unlike thoughts kept to oneself, conversations with AI are stored on company servers where automated systems and staff can read them, and they may bear on third parties' lives, so there must be an exception that is narrow and published in advance. Referral to law enforcement should occur only when a human has judged, from facts revealed in the conversation, that there is a serious risk to life or physical safety, as a rule only when the risk is imminent and the target identifiable; the information handed over should be minimized, and the state should separately control any use of a referred conversation to punish the user for that conversation itself.

The public documents analyzed did not yet meet these conditions. Even companies with narrow criteria had not withdrawn broad sharing provisions, and none disclosed the specific facts and exception conditions that decide a referral, a requirement of an identifiable target, or post-referral notification. In other words, the specific conditions proposed by this study were not disclosed in the companies' public documents. Reading each company's documents together also showed that, for several companies, the scope of referral cannot be read from any single document, so the relations among a company's documents are part of what it discloses. As conditions for justifying proactive referral, this study proposes specific criteria that limit corporate discretion together with advance disclosure of the procedure.

**Analyzed policy documents**

**Statutes and cases**

# Supplementary Material: When Is Disclosing Private Conversations with AI to the State Justified?

## Overview

This Supplementary Material supports the readings in the main text. S1 presents each company's readings together with the source texts on which they are based; S2, the register of analyzed documents; S3, the SHA-256 values of the saved source files; S4, the judgment categories for each criterion, each evaluator's readings, supplementary element-level summaries, and alternative readings that could change the judgments; S5, the record of the evaluation process; and S6, supplementary tables. Document IDs (O1–X3) are as in Table S6-3.

## S1. Readings by Company and Source Texts

For each company, the reading for each criterion and the reasons for it are given first, followed by the passages on which it is based, reproduced verbatim in the original language. Principle numbers refer to the four reading principles in Section V.4 of the main text. Criterion 7 (separate control over subsequent use) is a condition for the state and was therefore excluded from the company evaluations; it is recorded only under "For reference." Where a cell contains several passages, they are joined with " … ", and quotations are often parts of sentences. The location of each quoted segment is noted in the Description column.

### S1-1. OpenAI

**Severity.** The 2026 notice dealing specifically with referral (O6) states that the company refers cases involving "an imminent and credible risk of harm to others" and immediately afterward explains this criterion as a risk of "planned violence," thereby narrowing referral to risks to life or physical safety (Principle 2). The privacy policy updated later (O5) retained sharing with governments to protect rights and property, to address violations of the terms or the law, and to protect the safety of users and the public; an alternative reading that treats this as the scope of referral is given in Section VI.4 of the main text.

**Imminence.** Both notices state "imminent" as a condition for referral (O1, O6). The privacy policy (O5) permits sharing with governments without imminence, but because it is a general provision, the referral-specific provision was followed (Principle 2).

**Identifiable target.** The safety notice narrowed referral to risks to others, but by stating that a referral may be made even when the target, means, and timing are not explicitly discussed, it did not go so far as to require that the target be apparent (O6).

**Judgment.** The company stated that trained personnel assess flagged activity in context, looking at the content of the interaction, the surrounding conversation, and patterns of behavior over time, thereby specifying who decides and on what material, but it did not state what facts must emerge for a referral.

**Recipient.** The company described referral to law enforcement as applying to risks to others and stated that it is currently not referring self-harm cases to law enforcement (O1). Because the 2026 notice did not change this, it was treated as still in effect (Principle 1).

**Information minimization.** The safety notice did not state what information is transferred. The privacy policy (O5), which supplies this, states that Personal Data "including information about your interaction with our Services" may be shared with government authorities and others, thereby disclosing the types, but says nothing about limiting the amount (Principle 2).

**Advance notice.** The company disclosed the sequence of detection, review by personnel, in-depth investigation, and referral, together with the referral criteria (O6).

**Post-referral notification.** Notification after a proactive referral was not stated. For government requests, the company stated that it provides prior notice and delayed notice (O2).

### Table S1-1. OpenAI: Source Texts

| Document | Description | Original text (English) |
|---|---|---|
| O6 | Detection and review | “We use automated detection systems to identify potentially concerning activity at scale.” … “When an account or conversation is flagged, it is assessed in context by trained personnel.” … “Their role is to assess the flagged activity in context, including the content of the interaction, surrounding conversation, and any relevant patterns of behavior over time.” |
| O6 | In-depth investigation | “Where we assess that a case presents indicators of potentially serious, real-world harm, it is escalated for a more in-depth investigation, including assessing the overall level of risk using structured criteria.” |
| O6 | Referral requirement and recipient | “When conversations indicate an imminent and credible risk of harm to others, we notify law enforcement.” |
| O6 | Flexibility of referral criteria | “our referral criteria is flexible to account for the fact that a user may not explicitly discuss the target, means, and timing of planned violence in a ChatGPT conversation but that there may still be potential risk of imminent and credible violence” |
| O1 | Referral criteria in the earlier notice and handling of self-harm | “If human reviewers determine that a case involves an imminent threat of serious physical harm to others, we may refer it to law enforcement” … “We are currently not referring self-harm cases to law enforcement” |
| O5 | Sharing provision | “We may share your Personal Data, including information about your interaction with our Services, with government authorities, industry peers, or other third parties in compliance with the law” … “(ii) to protect and defend our rights or property, (iii) if we determine, in our sole discretion, that there is a violation of our terms, policies, or the law” … “to protect the safety, security, and integrity of our products, employees, users, or the public” |
| O7 | Specificity of the alert pathway (Section VII.1): Trusted Contact | “If our automated monitoring systems detect the user may be talking about self-harm in a way that indicates a serious safety concern, ChatGPT lets the user know that we may notify their Trusted Contact” … “From there, a small team of specially trained people review the situation.” … “every notification undergoes trained human review before it is sent, and we strive to review these safety notifications in under one hour” … “It does not include chat details or transcripts to protect user privacy.” |
| O2 | User notice of government requests (contrast in Section VI.3) | “OpenAI provides notice to the user before disclosing the information, unless prohibited by law.” … “OpenAI provides delayed notice to users after disclosure where no longer prohibited by law from doing so” |

## S1-2. Meta

**Severity.** The sharing provision (Me3) lists “Emergency situations, such as risk of death or imminent bodily harm” as one of seven situations in which information is shared, but because the same list also includes protecting rights and property and preventing fraud, it was read as describing risk broadly together with property and rights.

**Imminence.** Imminent bodily harm is an example of an emergency situation in the list, and information may also be shared in other situations, so referral was not narrowed to imminent cases (Principle 3).

**Identifiable target.** The same list includes “Protecting you or others,” naming both the user and others.

**Judgment.** The Meta AI Terms (Me6) state that review of interactions with AIs may be automated or manual (human), but a statement that a human may look at them does not mean that a human decides on referral. The terms of use (Me5) state that the company employs dedicated teams, develops detection systems, and may contact law enforcement “based on our assessment,” but do not state who makes that assessment (an alternative reading that treats this as a judgment by the dedicated teams is in S4). The sharing provision states only a “good faith belief.”

**Recipient.** The company named law enforcement, government authorities, affiliates, and others as recipients, but did not limit sharing with law enforcement to risks to others.

**Information minimization.** Neither the types nor the amount of information transferred was stated.

**Advance notice.** By disclosing the situations in which information is shared (Me3) and the detection of harmful conduct and contact with law enforcement “based on our assessment” (Me5), the company disclosed that detection leads to referral, but did not disclose who makes the assessment. Me6 links the two documents by stating that interactions with AIs are used in accordance with the Meta Terms of Service and Privacy Policy.

**Post-referral notification.** Notification after a proactive referral was not stated. For government requests, the company stated that it provides notice and delayed notice (Me1).

**(For reference) Subsequent use.** By listing “Investigating suspicious activity” among the situations in which information is shared with law enforcement, the company presupposed that transferred information is used in investigations.

### Table S1-2. Meta: Source Texts

| Document | Description | Original text (English) |
|---|---|---|
| Me3 | Sharing provision (full list of situations); segments: Section title … List | “How we promote safety, security and integrity” … “We share your information with law enforcement, government authorities, Meta Companies and other third parties (including industry peers) when we have a good faith belief it’s necessary to detect, prevent and address a variety of situations, such as: Unauthorized use of our Products / Violations of our terms and policies / Investigating suspicious activity / Protecting ourselves, including our rights, property, personnel or Products / Preventing abuse, fraud, or other harmful or illegal activity, on and off our Products / Protecting you or others, including as part of investigations or regulatory inquiries / Emergency situations, such as risk of death or imminent bodily harm” |
| Me3 | Users without an account | “We can take action according to our Community Standards against someone without an account interacting with Meta Products. For example, we can share information with law enforcement when we believe there is a genuine risk of death or imminent bodily harm.” |
| Me3 | Detection and review | “When our algorithms detect that someone might need help, a reviewer can review their post and offer support if needed.” |
| Me5 | Detection, assessment, and contact with law enforcement | “We employ dedicated teams around the world” … “develop advanced technical systems to detect potential misuse of our Products, harmful conduct towards others, and situations where we may be able to help support or protect our community” … “If we learn of content or conduct like this, we may take appropriate action based on our assessment that may include - notifying you, offering help, removing content, removing or restricting access to certain features, disabling an account, or contacting law enforcement.” |
| Me6 | Review and links between documents | “In some cases, Meta will review your interactions with AIs, including the content of your conversations with or messages to AIs, and this review may be automated or manual (human).” … “Meta uses Content and related information as described in the Meta Terms of Service and Meta Privacy Policy” … “To monitor your use of AIs for compliance with these Terms and applicable laws and to report violations of applicable laws or regulations as required by law” |
| Me7 | Emergency disclosure (government request context) | “In emergencies, law enforcement may submit requests without legal process. Based on the circumstances, we may voluntarily disclose information to law enforcement where we have a good faith reason to believe that the matter involves imminent risk of serious physical injury or death.” |
| Me4 | Specificity of the alert pathway (Section VII.1): parental alerts | “We worked with parents and experts to understand which AI conversations warrant an alert — such as those where a teen makes a clear reference to hurting themselves, even if that reference is subtle. We then built a dedicated AI system to identify these conversations.” … “as we continue to improve our detection, all chats flagged by our AI will be manually reviewed before an alert is sent” … “If a teen’s intent is ambiguous, we’ll err on the side of caution and alert the parent.” |
| Me1 | User notice and delayed notice of government requests (contrast in Section VI.3) | “Our policy is to notify people who use our service of requests for their information prior to disclosure unless we are prohibited by law from doing so or in exceptional circumstances, such as child exploitation cases, emergencies or when notice would be counterproductive.” … “We will also provide delayed notice upon expiration of a specific non-disclosure period in a court order and where we have a good faith belief that exceptional circumstances no longer exist” |

## S1-3. Microsoft

**Scope of documents.** The law enforcement documents (M1, M2) apply company-wide, and the privacy policy (M3) applies to Copilot. The Q&A in M1 on what happens “absent a formal legal request” addresses unrequested disclosure and was therefore read as a basis (Principle 4). An alternative reading that does not apply this Q&A to Copilot is in S4. The provisions in the same answer concerning forums, Xbox, and reports from customers fall outside the definition of proactive referral, because they are triggered by public postings or reports.

**Severity.** The Q&A on unrequested disclosure (M1) singles out emergencies involving death or serious physical injury, but the same answer also provides for reporting crimes against the company, its employees, or its customers, and therefore does not narrow referral to life or physical safety.

**Imminence.** The Q&A (M1) names emergencies but also provides for reporting crimes, which does not require imminence.

**Identifiable target.** Targets are stated broadly, as customers, employees, the company, and “a person” (M1).

**Judgment.** Only the degree of belief is stated, as in “good faith belief” (M3) and “where we believe” (M1).

**Recipient.** The Q&A names law enforcement as the recipient but does not limit referral to risks to others. M3 does not state the recipient of disclosures to protect safety.

**Information minimization.** The Q&A (M1) limits the amount by stating that “limited information” is reported both in reporting crimes and in disclosures in emergencies, and the types of information that the Q&A does not state were supplied by the personal data in M3 “including your content (like the content of your emails in Outlook.com or files in OneDrive)” (Principle 2). “Limited” does not mean the same as the minimum necessary, and its standard was not stated.

**Advance notice.** Only the situations in which referral occurs (emergencies) were disclosed.

**Post-referral notification.** Notification after a proactive referral was not stated. For government requests, the company stated that it provides prior notice and delayed notice (M1).

### Table S1-3. Microsoft: Source Texts

| Document | Description | Original text (English) |
|---|---|---|
| M3 | Disclosure provision and information type | “we retain, access, transfer, disclose, and preserve personal data, including your content (like the content of your emails in Outlook.com or files in OneDrive), when we have a good faith belief that it is necessary to do any of the following” … “Protect the safety of our customers, organizations, and the public” |
| M3 | Rights and property provision | “Protect the rights or property of Microsoft, including enforcing the terms governing the use of our services. If we receive information indicating our services are used to traffic in stolen property, we may refer the matter to law enforcement.” |
| M1, M2 | Forum, Xbox, and customer-report provisions; segments: Q&A question | “Does Microsoft provide any data to governments absent a formal legal request?” … “On occasion, we also report some limited information about a user when we have reason to believe the individual is about to harm themselves or someone else due to a public posting on one of our forums, on Xbox LIVE, or through referrals from other customers.” |
| M1, M2 | Crime victim provision | “If one of our customers or employees, or Microsoft itself, is the victim of a crime, we may report some limited information to law enforcement.” |
| M1, M2 | Emergency disclosure without a request; segments: Q&A question … From the answer | “Does Microsoft provide any data to governments absent a formal legal request?” … “Additionally, consistent with applicable law and industry practice, Microsoft sometimes discloses limited information to law enforcement where we believe the disclosure is necessary to prevent an emergency involving danger of death or serious physical injury to a person.” |
| M1 | User notice of government requests (contrast in Section VI.3) | “Microsoft gives prior notice to users whose data is sought by a law enforcement agency or other governmental entity, except where prohibited by law.” |

## S1-4. Anthropic

**Severity.** The privacy policy (A6) states "serious harm to any person or to property," and the terms of use (A7) state reporting "at our sole discretion." The reference to "imminent physical harm or death" in the government request policy (A1) is an exception to the principle that information is disclosed in response to government requests only pursuant to valid legal process, so it was treated as disclosure in response to emergency government requests and excluded as a basis (Principle 4). An alternative reading that takes this sentence to include unrequested disclosure as well is in S4.

**Imminence.** A6 and A7 contain no imminence condition. The emergency in A1 appears in the context of government requests and was not counted.

**Identifiable target.** Targets are stated broadly, as "any person."

**Judgment.** A6 states only a "good-faith belief," and A7 only sole discretion. The detection sentence stating that systems flag inputs or outputs (A6) was not linked to the referral decision.

**Recipient.** Government authorities, law enforcement, and others (A6) and law enforcement (A7) are named, but referral is not limited to risks to others. A8 describes pointing users who express suicidal or self-harm thoughts toward helplines and other human support but does not state whether such conversations are excluded from referral to law enforcement, so it did not change this reading.

**Information minimization.** By describing the information as "including but not limited to Inputs, Outputs, or Actions" (A7), the company disclosed the types but left the scope open.

**Advance notice.** The privacy policy (A6) disclosed the situations in which referral occurs by naming sharing to prevent "serious harm to any person or to property," but did not disclose how detection and review are linked to referral.

**Post-referral notification.** Notification after a proactive referral was not stated. For government requests, the company stated that it provides notice (A1).

### Table S1-4. Anthropic: Source Texts

| Document | Description | Original text (English) |
|---|---|---|
| A6 | Sharing provision | "We may share personal data with government authorities, law enforcement, or other third parties where, based on the information available to us, we have a good-faith belief that disclosure is reasonably necessary to" … "prevent serious harm to any person or to property" |
| A6 | Detection | "If our systems flag Inputs or Outputs for harmful content or for potentially violating our policies" |
| A7 | Referral at sole discretion | "We reserve the right, at our sole discretion, to report information from or about you, including but not limited to Inputs, Outputs, or Actions to law enforcement." |
| A8 | Response to suicidal or self-harm thoughts (Criterion 5) | "If someone expresses personal struggles with suicidal or self-harm thoughts, Claude should react with care and compassion while pointing users towards human support where possible: to helplines, to mental health professionals, or to trusted friends or family." |
| A1 | Emergency exception (government request context); "this policy" = disclosure in response to government requests only pursuant to valid legal process | "We will make an exception to this policy if we believe there is an emergency that may result in imminent physical harm or death, and providing such information without delay may avert that emergency." |
| A1 | User notice of government requests (contrast in Section VI.3) | "We will also provide users notice if their data is requested, unless we believe we're legally prohibited from doing so, or other rare exceptions apply, including where a child is at risk of harm or in cases of emergency." |

## S1-5. xAI

**Scope of documents.** The privacy policy analyzed (X2) applies to the Grok apps and Grok.com and states that use of Grok on the X platform is governed by X's policies.

**Severity.** Protecting "the personal safety of SpaceXAI, its customers, or any person" is listed in the same provision as protecting rights and property.

**Imminence.** Not stated.

**Identifiable target.** Targets are stated broadly, as "any person."

**Judgment.** The disclosure provision states neither who decides nor any degree of belief. Review by personnel (X3) is described as review for business purposes such as investigating security incidents and is not linked to referral.

**Recipient.** The policy states only that information may be disclosed to "others."

**Information minimization.** The policy describes "User Content" (inputs and outputs) separately and states that it may be disclosed "for legal purposes," thereby disclosing the types, but says nothing about limiting the amount. The legal purposes include protecting the personal safety of any person.

**Advance notice.** Only the purposes were disclosed.

**Post-referral notification.** Not stated.

### Table S1-5. xAI: Source Texts

| Document | Description | Original text (English) |
|---|---|---|
| X2 | Disclosure provision (full context) and scope of application; segments: Section 4 title … Introductory sentence … Legal purposes … Earlier in the document, scope of application | "4. How we may disclose personal information" … "We may disclose your personal information to others." … "For legal purposes: To (i) comply with laws or to respond to lawful requests and legal process, (ii) protect the rights and property of SpaceXAI and our agents, customers, and others, including to enforce our agreements, policies, and terms of service, (iii) to protect against legal liability, or (iv) to protect the personal safety of SpaceXAI, its customers, or any person." … "For example, your use of X (previously Twitter), including use of Grok on the X platform, is governed by the X Privacy Policy and X Terms, not this SpaceXAI Privacy Policy." |
| X2 | User Content (information type) | "User Content: You may provide personal information in prompts and other content you input, such as files, images, audio, voice, video, and other material ("Input"). Outputs of the Service ("Output"), including responses Grok generates, are based on your Input (together, "User Content")." … "How we may disclose it: To our contracted service providers; in connection with business transfers, for legal purposes; and to our related companies." |
| X3 | Review | "Our authorized personnel may review how you use the Service and your User Content for specific business purposes, including improving product features, investigating security incidents and potential misuse of our Service, and complying with our legal obligations." |

## S1-6. Google

**Severity.** The general privacy policy (G4) stated that Google shares information when necessary to protect against harm to the rights, property, or safety of Google, its users, or the public. The sentence on death or serious physical harm in the government request document (G2) appears under the heading "Requests for information in emergencies" and was therefore read as disclosure in response to requests (Principle 3).

**Imminence.** Not stated.

**Identifiable target.** Named Google, its users, and the public.

**Judgment.** Stated only "good-faith belief." The Gemini notice (G1) stated that human reviewers review some data but did not link this to referral.

**Recipient.** Stated only that it shares information "outside of Google."

**Information minimization.** Stated only "personal information."

**Advance notice.** Only the purposes were disclosed.

**Post-referral notification.** Did not state notification after proactive referral. For government requests, it stated notice and delayed notice (G2).

### Table S1-6. Google: Source Texts

| Document | Description | Original text (English) |
|---|---|---|
| G4 | Sharing provision | "We will share personal information outside of Google if we have a good-faith belief that disclosure of the information is reasonably necessary to:" … "Protect against harm to the rights, property or safety of Google, our users, or the public." |
| G1 | Links between documents and human review; "these purposes" = the purposes of this section, such as providing, improving, and protecting the services | "This notice supplements the Google Privacy Policy" … "Human reviewers (including trained reviewers from our service providers) review some of the data we collect for these purposes." |
| G2 | Emergency disclosure sentence and its location (Principle 3); segments: Paragraph heading | "Requests for information in emergencies" … "If we reasonably believe that we can prevent someone from dying or from suffering serious physical harm, we may provide information to a government agency — for example, in the case of bomb threats, school shootings, kidnappings, suicide prevention, and missing persons cases." … "We still consider these requests in light of applicable laws and our policies." |
| G2 | User notice and delayed notice of government requests (contrast in Section VI.3) | "When we receive a request from a government agency, we send an email to the user account before disclosing information." … "And we might not give notice in the case of emergencies, such as threats to a child's safety or threats to someone's life, in which case we'll provide notice if we learn that the emergency has passed." |

## S2. Document Register

Table S2-1 lists the original titles, URLs, and scope of analysis of the 27 analyzed documents. Versions and access dates are given in Table S6-3. xAI's privacy policy (X2) was updated on October 11, 2026, after it was accessed, but the provisions used in the reading and the scope of application did not change.

### Table S2-1. Original titles, URLs, and scope of analysis of the analyzed documents

| No. | Original title | URL | Scope of analysis |
|---|---|---|---|
| O1 | Helping people when they need it most | openai.com/index/helping-people-when-they-need-it-most/ | Full text |
| O2 | Government User Data Request Policy (v2025.12) | cdn.openai.com/pdf/openai-law-enforcement-policy-v.2025-12.pdf | Full text |
| O4 | Report on Government Requests for User Data (July–December 2025) | cdn.openai.com/trust-and-transparency/report-2025h2-government-requests-for-user-data.pdf | Full text |
| O5 | US Privacy Policy | openai.com/policies/us-privacy-policy/ | Full text |
| O6 | Our commitment to community safety | openai.com/index/our-commitment-to-community-safety/ | Main text |
| O7 | Introducing Trusted Contact in ChatGPT | openai.com/index/introducing-trusted-contact-in-chatgpt/ | Main text |

| No. | Original title | URL | Scope of analysis |
|---|---|---|---|
| A1 | What is Anthropic's policy for handling governmental requests for user information? | support.claude.com/en/articles/9519291 | Full text |
| A2 | Law Enforcement Requests | support.claude.com/en/articles/9035075 | Full text |
| A5 | Government Requests Report (July–December 2025) | www-cdn.anthropic.com/ 5d453bc3285b8e0c101b7193b5765419920cee24.pdf | Full text |
| A6 | Privacy Policy | anthropic.com/legal/privacy | Full text (English version) |
| A7 | Consumer Terms of Service | anthropic.com/legal/consumer-terms | Preamble, Sections 8 and 12 |
| A8 | Protecting the wellbeing of our users | anthropic.com/news/protecting-well-being-of-users | Paragraph on responding to self-harm |
| G1 | Gemini Apps Privacy Hub | support.google.com/gemini/answer/13594961 | Full text (including collapsed items) |
| G2 | How Google handles government requests for user information | policies.google.com/terms/information-requests | Full text |
| G4 | Google Privacy Policy | policies.google.com/privacy | "Sharing your information" section |
| G5 | Transparency Report: Global requests for user information | transparencyreport.google.com/user-data/ overview | Overview page (excluding country-specific and downloadable data) |
| M1 | Government Requests for Customer Data Report | microsoft.com/en-us/corporate-responsibility/ reports/government-requests/customer-data | On-page Q&A and chart alternative text |
| M2 | About our practices and your data | blogs.microsoft.com/datalaw/our-practices/ | Full text (including Q&A) |
| M3 | Microsoft Privacy Statement | microsoft.com/en-us/privacy/privacystatement | Common provisions and the "Artificial Intelligence and Copilot capabilities" provisions |
| Me1 | Information for law enforcement authorities | meta.com/safety/communities/law/guidelines/ | Full text |
| Me3 | Privacy Policy | facebook.com/privacy/policy/ | Main text and explanations (excluding the appendix on legal bases for processing) |
| Me4 | Alerting Parents if Teens Show Signs of Distress in Conversations With Meta AI | about.fb.com/news/2026/07/keeping-parents-informed-teens-distress-conversations-meta-ai/ | Main text |
| Me5 | Terms of Service | facebook.com/terms.php | Section 1.5 |
| Me6 | Meta AIs Terms of Service (version for regions other than the UK, Brazil, and the EU) | facebook.com/legal/ai-terms | Sentences on review and use |
| Me7 | Government Requests for User Data | transparency.meta.com/reports/government-data-requests/ | Global and U.S. pages; two Q&A items |
| X2 | Privacy Policy | x.ai/legal/privacy-policy | Full text |

| No. | Original title | URL | Scope of analysis |
|---|---|---|---|
| X3 | Terms of Service - Consumer | x.ai/legal/terms-of-service | License and review provisions |

**Table S2-2. Documents excluded after collection (6)**

| No. | Document | Reason for exclusion |
|---|---|---|
| O3 | OpenAI government request report (second half of 2024) | Only each company's most recent report was analyzed (O4) |
| A3, A4 | Anthropic government request reports (second half of 2024; first half of 2025) | Same reason (A5) |
| G3 | Google Transparency Report help page | A help page explaining the report's figures; not one of the five document types |
| Me2 | KWTX news report (Meta's parental alerts, July 16, 2026) | A news report, not a company document; replaced by the original Meta Newsroom post (Me4) |
| X1 | Earlier version of the xAI Privacy Policy (effective July 10, 2025) | Only the version current at the time of access (X2) was analyzed |

**Table S2-3. Documents reviewed but not included**

| Document | Version | Reason for non-inclusion |
|---|---|---|
| OpenAI Terms of Use | Effective January 1, 2026 | No conditions for passing user information to law enforcement or other authorities |
| OpenAI Usage Policies | Revised October 29, 2025 | Usage policies are not one of the analyzed document types; external reporting is mentioned only for child sexual abuse material |
| Google Terms of Service | Effective July 30, 2026 | No conditions |
| Microsoft Services Agreement; Copilot supplemental terms | Effective September 30, 2026; effective August 18, 2026 | No conditions (mentions only automated and human review) |
| Anthropic Usage Policy | Version scheduled to take effect November 12, 2026 | A usage policy; mentions detection and monitoring but sets no conditions for external reporting |
| Meta Korea Supplemental Privacy Policy | Applicable from September 23, 2026 | Covers the exercise of rights, retention periods, and security measures; no criteria for provision of information to law enforcement |
| Updated version of the xAI Privacy Policy | Updated October 11, 2026 | The version following X2; the disclosure provisions and scope of application used in the evaluation are the same (only the short name, dates, the paragraph on payment information, and one sentence differ) |

## S3. SHA-256 Values of the Saved Source Files

The values below were computed on October 12, 2026, for the text files saved at the time of access; hashes of the original screens as displayed at the time of access or of the PDF files themselves were not recorded. There are 28 files, consisting of the texts of the 27 documents and 1 alternative text for the M1 chart. These files are available from the corresponding author on request, and their identity can be checked against the values below.

**Table S3. SHA-256 of the saved files**

| File | Size (bytes) | SHA-256 |
|---|---|---|
| A1.txt | 1284 | 3acf5dd569b4b01c77915ef2f04adaab0079f4afe8f0e213dcc491d0b3cf7f54 |
| A2.txt | 1187 | 4c4992f23a561ec17db4c46456e375bb3fb9a21dd540584339d1ed1baa6d23bb |
| A5.txt | 3526 | 49dda87ab4f46d4dd585d4709e8021ecb8c9d61bde43a7ccb8bc25cc2757eb94 |
| A6.txt | 40509 | 9d818008e687febabdab640bc92cea38de7350e807ebbeb062f3a5beae41ee75 |
| G1.txt | 57202 | c4455f88a2ae5bf20caff262633e7de80a89df26ee8cfc2da494ec0b0bb16f18 |
| G2.txt | 6414 | 1a5de06a483f13c3aa6ea28f49cc00c479fc8388568c6c6d02ba53009390b54e |
| M1.txt | 20570 | fb141109e5dc70c952a8b35bc6d494fa21fbea353084ee7b13c968c2c8c8c258 |
| M2.txt | 26213 | fe072456371df456ebaca1731859a98f14c1985eb6fbd17fccbfc5adcb48a730 |
| M3.txt | 69701 | 69da38356717a52bb96e86508cfb0280e950680cb836ccedbdd411a5838c3bf7 |
| Me1.txt | 11229 | 714423a43917ca5f674fd3292df1da737e4c83f26e0378198642e7c89a76ecbd |
| Me3.txt | 93743 | 6328bf8e1e16ff934396e8701c659ae8eedfd61f32aa23817e41aaacbb7ddf15 |
| O1.txt | 8969 | d50f61383dab3b861bc633ef78670bf1e47736f640aa4aeda285b52911be241b |
| O2.txt | 9387 | a1cc010b6ac24887bc872e84ec1161369ad100990f3bf5e483fa83acc61158b0 |
| O4.txt | 2357 | 81d894f3666e9e8eb9f600b79f4d915d5d9e1fbd656463c4ef1c1f5d8fb5d3d6 |
| O5.txt | 29174 | 084bc6d82118cd2df0083d6c8e7fb0cd06e89472cf9a6d30db2018c774b034e3 |
| X2.txt | 27705 | f7636b80c2dbf7231a5e5a46985c4c97ff2fafa6455dca3589e6c1ab211f45c4 |
| A7.txt | 2369 | 6e504ad8f979b24d6e7b44c0fa2733070550fde95116e3c3d0c0c4611eea6b40 |
| A8.txt | 1622 | a1405086564be5c04e48a7ef5046c07b5d69769cf571ffd781e01e44fb658db0 |
| G4.txt | 1188 | 530821e4dbfbf8ba540e367e3341138e1c566f838efa8673343fe33ac3140be0 |
| G5.txt | 1290 | e8ae05d491b517c0358c9b2d1f2d3ba7de318594ec16c8f4d493fc514ae5ff78 |
| M1_chart.txt | 1059 | a7efc96fd0c655fbff0367ea6b8dbd7fd400a36e35bcb1288a31b733d6857d10 |
| Me4.txt | 4236 | 4b9909f38ff3b06da5c079cec6d46d8ae3d0d91c7442024b079b8934e1464fb0 |
| Me5.txt | 1832 | df1c35c6bb0a9dc5ab8c7a6c08f5485b2d14bd1db96cce43240539597fb55329 |
| Me6.txt | 1517 | c1c8c68034274dddef6fb3305ce57af0d244de3546d47cd77c3531b2506d7df1 |
| Me7.txt | 2754 | 25e1deecb563821798f711f634a11a0fa7ee3c2b775e02ae70f30f5e0859f002 |
| O6.txt | 8579 | 975fa3c589d10c667098e4005345f9be71172de2df7aee78c00ab92f9365b4d5 |
| O7.txt | 4014 | ecdd652166251c0617cfd73a384adb0872a081ee7f3867b68cc31de3a5b8ee11 |
| X3.txt | 2424 | 63e2de3e91420518e2eaa1a172798802ab1f97ace11c29b2d61fbb5b9d87e45c |

## S4. Judgment Categories, Evaluators' Readings, and Alternative Readings

Three evaluators (A, B, and C), all of whom are authors of this study, independently read each company's documents for Criteria 1–6, advance notice, and post-referral notification, following the reading principles in Section V.4 of the main text and the judgment categories below, without seeing one another's readings. The evaluators were given the scope rules (disclosure in response to government requests, the alert pathway, statutory reporting of child sexual abuse material, and referrals triggered by public posts or tips are not evaluated), the reading principles, guidelines setting out the judgment categories and their boundaries, and the source texts of the analyzed documents. The evaluators first read the 48 cells under a principle that treated broad provisions as the scope the company had retained, and after Principle 2 was set to give precedence to referral-specific provisions, they re-read the 7 cells it affected in the same manner (Table S4-3). Of the final 48 cells, the three evaluators' readings agreed on 41, and there was no cell on which all three read differently. In pairwise comparisons, A and C agreed on 44 cells, and A and B, and B and C, each agreed on 43 cells. In the first round, before re-reading, all three readings agreed on 40 cells (83%), and all 8 divergent cells were split 2–1; pairwise, A and C agreed on 43 cells, A and B on 44, and B and C on 41. The changes in agreement between the first round and the final judgments arose in the three re-read Microsoft cells (severity, imminence, and information minimization) (Table S4-3). By criterion, the cells on which all three evaluators agreed numbered 6 each for judgment, recipient, imminence, identifiable target, and post-referral notification; 4 each for severity and advance notice; and 3 for information minimization. The 7 divergent cells followed the reading shared by two or more evaluators. Of these, 2 cells, severity and information minimization for Microsoft, are cases in which a company described the scope of referral differently across documents or provisions (Section VII.1 of the main text). Element-level readings for the multi-element criteria are given in Table S4-4.

**Table S4-1. Judgment categories by criterion**

| Criterion | Judgment categories (from most specific) |
|---|---|
| 1. Severity | Narrowed referral to risks to life or physical safety (including violence). / Mentioned risks to life or physical safety as grounds for referral but did not narrow referral to them. / Described risk broadly, e.g., as “safety” or “harm” (listing property and rights together). / Did not state the type of risk. |
| 2. Imminence | Narrowed referral to risks that are about to occur. / Mentioned imminent or emergency situations but did not narrow referral to them. / Did not mention imminence. |
| 3. Identifiable target | Stated that referral occurs only when the target of the risk (person, place, group) is identifiable. / Narrowed referral to risks to others but did not go so far as to require that the target be identifiable. / Described the target broadly, as both oneself and others, or as “any person” or “the public.” / Did not state who is at risk. |
| 4. Who decides, and on what basis | Stated that a person reviews and what facts must emerge for a referral. / Stated that a person reviews and what the judgment is based on (conversation content, surrounding context, etc.), but not what facts must emerge for a referral. / Stated only that a person reviews. / Did not mention human review and stated only a degree of belief (“when we have a good-faith belief,” “suspicion,” etc.). / Stated neither who decides nor on what basis (“when necessary,” discretion only). (This category combines (i) the basis for judgment, (ii) human review, and (iii) disclosure of the decisive factors in the main text; element-level judgments are in Table S4-4.) |
| 5. Appropriate recipient | Stated that law enforcement referral is limited to risks to others and that self-harm is not sent to law enforcement. / Stated that risks to others are reported to law enforcement but did not mention self-harm, so it is unclear whether self-harm may also go to law enforcement. / Identified the recipient (law enforcement, government agencies, etc.) but did not limit law enforcement referral to risks to others. / Did not identify the recipient (only “disclose” or “third parties”). |
| 6. Information minimization | Identified the types of information passed on and stated that the amount is limited.* / Identified only the types of information passed on (e.g., “IP addresses, etc.,” “content such as the content of emails”), with nothing on reducing the amount. / Did not say what information, stating only that the amount is reduced (e.g., “report some limited information”). / Did not say what or how much is passed on (only “personal information” or “information”). |
| 8. User notice and transparency — Advance notice | Disclosed both the procedure from detection to referral (who, in what order) and the referral criteria. / Disclosed the referral criteria and that detection and review lead to referral, but not who decides. / Disclosed only the situations in which referral occurs (criteria), not the link between detection and review and referral. / Disclosed only the purpose (“to protect safety”) or discretion; even where detection or review was described, it was not linked to referral. |
| 8. User notice and transparency — Post-referral notification | Stated that the user is notified after a proactive referral. / Did not mention notification after a proactive referral. |

“Detection and review” in advance notice refers to activities that examine users’ conversations or usage to find and confirm risks or violations. It covers cases in which a document states procedures for examining conversations, such as automated detection systems, review personnel, or screening of inputs and outputs; a provision granting authority alone, such as “may determine (find) a violation,” was not treated as detection or review. This rule was applied equally to all companies.

* In the evaluation materials, this category was worded “stated that only what is necessary is transferred,” but because the guidelines included under “amount” not only “only what is necessary” but any wording that reduces the amount transferred, such as “some” or “limited,” it has been reworded to match the actual evaluation. This category does not mean that a standard of the minimum necessary for the purpose was disclosed. From Table S4-2 onward, it is abbreviated as “types and limit on amount.”

**Table S4-2. Cells on which evaluators’ readings diverged**

| Company | Criterion | Evaluator A | Evaluator B | Evaluator C | Final |
|---|---|---|---|---|---|
| Meta | Severity | Described broadly together with property or rights | Mentioned life or physical safety without narrowing to it | Described broadly together with property or rights | Described broadly together with property or rights |
| Meta | Information minimization | Neither types nor amount | Neither types nor amount | Types only | Neither types nor amount |
| Microsoft | Severity | Mentioned life or physical safety without narrowing to it | Described broadly together with property or rights | Mentioned life or physical safety without narrowing to it | Mentioned life or physical safety without narrowing to it |
| Microsoft | Information minimization | Types only | Types and limit on amount | Types and limit on amount | Types and limit on amount |
| Anthropic | Advance notice | Criteria (situations) only | Purpose or discretion only | Criteria (situations) only | Criteria (situations) only |
| xAI | Information minimization | Neither types nor amount | Types only | Types only | Types only |
| Google | Advance notice | Purpose or discretion only | Purpose or discretion only | Criteria (situations) only | Purpose or discretion only |

**Table S4-3. Cells re-read after the change to Principle 2**

| Company | Criterion | Initial reading | Re-reading | Dissenting evaluator |
|---|---|---|---|---|
| OpenAI | Severity | Mentioned life or physical safety without narrowing to it | Narrowed to life or physical safety | Initial: none / Re-reading: none |
| OpenAI | Imminence | Mentioned imminence or emergency without narrowing to it | Narrowed to imminent risks | Initial: none / Re-reading: none |
| OpenAI | Identifiable target | Target described broadly | Narrowed to risks to others | Initial: none / Re-reading: none |
| Microsoft | Severity | Described broadly together with property or rights | Mentioned life or physical safety without narrowing to it | Initial: C (Mentioned life or physical safety without narrowing to it) / Re-reading: B (Described broadly together with property or rights) |

| Company | Criterion | Initial reading | Re-reading | Dissenting evaluator |
|---|---|---|---|---|
| Microsoft | Imminence | Mentioned imminence or emergency without narrowing to it | Mentioned imminence or emergency without narrowing to it | Initial: B (Imminence not stated) / Re-reading: none |
| Microsoft | Identifiable target | Target described broadly | Target described broadly | Initial: none / Re-reading: none |
| Microsoft | Information minimization | Types only | Types and limit on amount | Initial: C (Types and limit on amount) / Re-reading: A (Types only) |

The table below is a supplementary element-level summary of the final categories for judgment (Criterion 4), information minimization (Criterion 6), and advance notice. Because the categories do not distinguish every combination of elements (e.g., disclosing only the types of information and disclosing only a limit on the amount are not nested levels), this table cannot fully reconstruct the presence of each element or the agreement among evaluators.

**Table S4-4. Supplementary element-level summary for multi-element criteria**

| Company | Human judgment | Materials for judgment | Referral threshold | Types of information | Amount of information | Referral situations | Link to detection and review | Procedure |
|---|---|---|---|---|---|---|---|---|
| OpenAI | Stated | Stated | – | Stated | – | Stated | Stated | Stated |
| Meta | – | – | – | – | – | Stated | Stated | – |
| Microsoft | – | – | – | Stated | Stated | Stated | – | – |
| Anthropic | – | – | – | Stated | – | Stated | – | – |
| xAI | – | – | – | Stated | – | – | – | – |
| Google | – | – | – | – | – | – | – | – |

### Alternative Readings

Besides the priority among documents discussed in Section VI.4 of the main text, the following three readings could change the judgments. Table 4 in the main text follows the reading adopted by this study for each item.

**Anthropic's emergency exception.** Anthropic's government request policy (A1) states the emergency exception as an exception to the principle that information is disclosed in response to government requests only pursuant to valid legal process. This study treated it as disclosure in response to emergency government requests and excluded it as a basis (Principle 4), but if it is read as a sentence that also covers notifying authorities first without a request, Anthropic becomes a company that cites life or physical safety and imminence as one case for referral.

**Scope of Microsoft's Q&A.** Microsoft's Q&A on what happens "absent a formal legal request" (M1) is a company-wide document, and the documents provide no basis for its applying to Copilot conversations. Without this Q&A, Microsoft drops out of the companies that cited life or physical safety and imminent risk, its statement that it limits information disappears, and its advance notice also shifts to disclosing purposes only.

**Meta's "based on our assessment."** There is also a reading in the opposite direction. If "based on our assessment" in Meta's Terms of Service (Me5) is read to mean that dedicated teams decide whether to refer, Meta also becomes a company that has stated that humans decide, and its advance notice moves closer to disclosing the procedure.

## S5. Record of the Evaluation Process

The evaluators were given the scope rules and reading principles, guidelines setting out the judgment categories and their boundaries, key passages and their translations for each company, and the full source texts of the analyzed documents.

After the materials were adjusted to the scope of evaluation, the three evaluators judged the 48 items once. After the interpretive principle on priority among documents was settled, the three evaluators re-judged only the 7 items it affected (Criteria 1–3 for OpenAI; Criteria 1–3 and 6 for Microsoft) and finalized the judgments, and all other items kept their first-round judgments (Table S4-3). The final agreement rate was calculated on the judgments reflecting this re-judgment.

## S6. Supplementary Tables

These tables supplement the analysis in the main text. Tables S6-1 and S6-2 supplement Sections VI and VII, and Table S6-5 supplements Section V; Tables S6-3 and S6-4 list the analyzed documents and the reported incidents.

**Table S6-1. Results by priority among documents**

| Criterion | Referral-specific provisions prevail (Principle 2) | General provisions prevail |
|---|---|---|
| 1. Severity | OpenAI narrowed referral to life or physical safety (planned violence). Microsoft mentioned life or physical safety separately but also provides for reporting crime victimization | None narrowed. OpenAI mentioned life or physical safety without narrowing to it; Microsoft described it broadly together with property or rights |
| 2. Imminence | OpenAI narrowed referral to imminent risks | None narrowed |
| 3. Identifiable target | OpenAI narrowed referral to risks to others (without stating whether the target must be identifiable from context) | None narrowed |
| 6. Information minimization | Microsoft identified the types and limited the amount with “limited information” (no minimum-necessary standard) | None stated that it limits the amount |
| Same result under both readings | None stated a requirement that the target be identifiable or the facts that serve as the threshold for judgment. OpenAI alone disclosed human judgment, the procedure from detection to referral, and the exclusion of self-harm. No policy on notification after proactive referral was found in the documents reviewed | Same |

**Table S6-2. Application of the criteria to boundary cases**

| Case | Assessment against the criteria | Conclusion |
|---|---|---|
| Fiction, role-play, emotional venting (e.g., pouring out anger and writing “I want to kill them” with no means, timing, or preparation) | There is an expression of harm to life or physical safety, but when a person views the surrounding context it is creative writing or emotional expression, and no target, means, or preparation is evident (Criteria 2–4) | No referral. If needed, user-directed measures such as pointing to counseling resources |
| Immediate threat to others (e.g., writing that one will harm a specific person tonight with a knife one has) | Life or physical safety (1), imminence (2), and target (3) are evident, and a person viewing the surrounding conversation judges it to be a concrete and credible threat in context (4) | Referral. Only the content of the threat and the necessary identifying information are passed on (6) |

| Case | Assessment against the criteria | Conclusion |
|---|---|---|
| Concrete preparations with unknown target and timing (e.g., writing that one bought a gun to carry out an attack, without having set a date or place) | Imminence (2) and an identifiable target (3) are not confirmed, but means and preparations linked to an intent to attack are concretely evident | As a narrow exception, referral is permissible if a person confirms from the surrounding context an intent to attack and a realistic possibility of carrying it out. These requirements are disclosed in advance (8) |
| Mix of self-harm and harm to others (e.g., writing that one will harm one's family and then kill oneself) | Risk to others is assessed under Criteria 1–4 and risk of self-harm separately under Criterion 5 | Risk to others is referred if it meets Criteria 1–4 or a narrow exception disclosed in advance; risk of self-harm is routed to crisis support. Information passed on is reduced to the part concerning the risk to others |

All documents are primary sources posted directly by the companies, and the on-screen text was saved on the access date (2026). The numbers were assigned at the collection stage, so the numbers of documents excluded after collection (O3, A3, A4, G3, Me2, X1) are not used. Original titles, URLs, and scope of analysis are given in Supplementary Material S2.

**Table S6-3. Analyzed documents (27)**

| No. | Document | Version | Accessed |
|---|---|---|---|
| O1 | OpenAI safety notice (crisis response) | Posted August 26, 2025 | Oct. 9, 2026 |
| O2 | OpenAI law enforcement policy | v2025.12, effective January 1, 2026 | Oct. 9, 2026 |
| O4 | OpenAI government request report | Second half of 2025 | Oct. 9, 2026 |
| O5 | OpenAI U.S. privacy policy | Updated September 10, 2026 | Oct. 11, 2026 |
| O6 | OpenAI safety notice (community safety) | Posted April 28, 2026 | Oct. 12, 2026 |
| O7 | OpenAI safety notice (Trusted Contact) | Posted May 7, 2026 | Oct. 12, 2026 |
| A1 | Anthropic policy on handling government requests (help center) | Page date March 17, 2026 | Oct. 9, 2026 |
| A2 | Anthropic guide to law enforcement requests (help center) | Page date March 17, 2026 | Oct. 9, 2026 |
| A5 | Anthropic government request report | Second half of 2025 | Oct. 9, 2026 |
| A6 | Anthropic privacy policy | Effective September 10, 2026 | Oct. 11, 2026 |
| A7 | Anthropic consumer terms of service | Effective October 8, 2025 | Oct. 12, 2026 |
| A8 | Anthropic news (user wellbeing) | Posted December 18, 2025 | Oct. 12, 2026 |
| G1 | Google Gemini Apps privacy information | Updated September 24, 2026 | Oct. 9, 2026 |
| G2 | Google's handling of government requests | Version current at access | Oct. 9, 2026 |
| G4 | Google privacy policy | Effective October 1, 2026 | Oct. 12, 2026 |
| G5 | Google Transparency Report (user information requests) | July–December 2025 | Oct. 12, 2026 |
| M1 | Microsoft government request report (law enforcement requests) | Second half of 2025 | Oct. 9 and Oct. 12, 2026 |
| M2 | Microsoft principles for responding to law enforcement | May 29, 2024 | Oct. 9, 2026 |

| No. | Document | Version | Accessed |
|---|---|---|---|
| M3 | Microsoft privacy statement | Updated September 2026 | Oct. 11, 2026 |
| Me1 | Meta information for law enforcement authorities | Version current at access | Oct. 9, 2026 |
| Me3 | Meta privacy policy | Effective July 23, 2026 | Oct. 11, 2026 |
| Me4 | Meta Newsroom (parental alerts) | Posted July 16, 2026 | Oct. 12, 2026 |
| Me5 | Meta terms of service | Effective January 1, 2025 | Oct. 12, 2026 |
| Me6 | Meta AI terms | Effective May 13, 2026 | Oct. 12, 2026 |
| Me7 | Meta government request report | July–December 2025 | Oct. 12, 2026 |
| X2 | xAI privacy policy | Effective August 24, 2026 | Oct. 9, 2026 |
| X3 | xAI consumer terms of service | Updated September 11, 2026 | Oct. 12, 2026 |

The incidents below are not a basis for the readings; they are used only as examples that raise the problem and show the tensions between criteria and when the criteria changed. Arrest and the filing of charges are distinct from indictment or conviction.

**Table S6-4. Reported incidents**

| No. | Incident (company, date) | Flow of the conversation and outcome | Level of verification (main sources) |
|---|---|---|---|
| C1 | Tumbler Ridge shooting, Canada (OpenAI; account suspended June 2025, shooting February 2026) | After detection and human review, the account was suspended, but authorities were not notified because the case was judged not to meet the referral criteria. Eight people killed. Criteria made more flexible (described in the February 2026 letter) | News reports, company letter (Winter, 2026; OpenAI, 2026; CP24, 2026) |
| C2 | Arrest of a Claude user in Florida (Anthropic; messages September 26–27, 2026, reported October 5, 2026) | Law enforcement notified after automated monitoring and human review. Arrest on a written-threats charge. Indictment and trial not confirmed | News reports (Dellinger, 2026; Krištopaitytė, 2026) |

Sources for Table S6-4 that are not listed in the references of the main text are as follows.

**Table S6-5. Companies analyzed and policies analyzed**

| Company (chatbot) | Entity and scope of the policies analyzed |
|---|---|
| OpenAI (ChatGPT) | Consumer services. U.S. version of the privacy policy |
| Anthropic (Claude) | Anthropic, PBC. Consumer services such as Claude.ai |
| Google (Gemini) | Gemini Apps (G1); Google services as a whole (G2, G4, G5) |
| Microsoft (Copilot) | All consumer products (M3); company-wide (M1, M2) |
| Meta (Meta AI) | Meta Platforms, Inc. Facebook, Instagram, Messenger, and Meta AI |
| xAI (Grok) | SpaceXAI LLC. Grok app and Grok.com (excluding Grok within X and the API) |